\documentclass[a4paper,preprint,amsmath,prl,showpacs]{revtex4-1}
\usepackage{graphicx,caption,amssymb}
\usepackage[show]{ed}

\begin{document}
\title{Generalised fractional diffusion equations for subdiffusion on arbitrarily growing domains}
\author{C. N. Angstmann}
\email{c.angstmann@unsw.edu.au}

\affiliation{School of Mathematics and Statistics, UNSW, Sydney NSW, 2052, Australia}
\author{B. I. Henry}
\email{b.henry@unsw.edu.au}
\affiliation{School of Mathematics and Statistics, UNSW, Sydney NSW, 2052, Australia}

\author{A. V. McGann}
\email{a.mcgann@unsw.edu.au}
\affiliation{School of Mathematics and Statistics, UNSW, Sydney NSW, 2052, Australia}
                  
\date{\today}
\begin{abstract}
Many physical phenomena occur on domains that grow in time. When the timescales of the phenomena and domain growth are comparable, models must include the dynamics of the domain.
A widespread intrinsically slow transport process is subdiffusion. Many models of subdiffusion include a history dependence. This greatly confounds efforts to incorporate domain growth. Here we derive the fractional partial differential equations that govern subdiffusion on a growing domain, based on a Continuous Time Random Walk. This requires the introduction of a new, comoving, fractional derivative.

\end{abstract}

\maketitle

A wide range of important physical phenomena involve transport on expanding, and contracting, domains. Fundamental examples include, the diffusion of proteins within growing cells, the interactions of cells in a growing organism, and diffusion in an expanding universe.  The governing equations for reaction diffusion on growing domains and related studies of pattern formation have been considered in a series of publications, see for example,  \cite{CGK1999,M2001,CM2001,BYE2010,WBGM2011,YBEM2012,SSMB2015,YAE2016}. Domain growth has been shown to be fundamentally important to the development of patterns \cite{KA1995}. Here we consider the problem of subdiffusive transport on a growing domain by constructing a continuous time random walk (CTRW) and limiting to a fractional order partial differential equation (PDE). 

Subdiffusion, which is characterised by a sub-linear power-law scaling in time of the mean squared displacement, is common in biological systems with traps and obstacles \cite{S2007}, such as diffusion of molecules in spiny nerve cells \cite{SWDA2006}, diffusion across potassium channels in membranes \cite{MNHMT2010,WSTK2011}, and diffusion of HIV virions in cervical mucous \cite{SACSA2013}. Subdiffusion is also present in other physical systems such as cosmic rays \cite{S2005}, porous media \cite{LB2003}, and volcanic earthquakes \cite{AS2017}.  The generalisation of canonical mathematical diffusion models to incorporate subdiffusive transport, such as, reaction-diffusion PDEs \cite{HLW2006,SSS2006, F2010,AYL2010,ADH2013mmnp}, and Fokker-Planck PDEs \cite{BMK2000,SK2006,HLS2010,ADH2013mmnp}, has proven non-trivial. In the work below we show that this is also true for subdiffusion on a growing domain.

There are different theoretical approaches that have been used to model subdiffusive transport. One of the more rigorous approaches is to derive the governing equations from the stochastic process of a CTRW \cite{MW1965}. The CTRW describes transport of particles on a mesoscopic scale in which particles wait for a time, governed by waiting time probability density, before randomly jumping, governed by a jump length probability density, to another location. If the jump length density is symmetric with a finite variance and the expected waiting time is convergent, then the CTRW limits to the standard diffusion PDE \cite{HA1995,MK2000}. If the waiting time density is replaced with a heavy tailed power-law waiting time density, then the CTRW limits to a time subdiffusion fractional diffusion PDE \cite{HA1995,MK2000}.

In the following we start with the underlying stochastic process of a CTRW to derive master equations for subdiffusive transport on a growing domain.  In our derivation we first consider a mapping between the position $y$ on the growing domain at any time $t$ and a corresponding position $x$ on the original domain at time $t=0$. With this mapping we then transform the CTRW from the coordinates on the growing domain to a non-growing fixed domain. An auxiliary master equation for the evolution of the density on the fixed domain is derived. The auxiliary master equation is constructed so that the value of the density at a given $x$ and $t$ equates to the probability density on the growing domain for $y$ and $t$.
The diffusion limit of the master equation is taken to produce a fractional diffusion equations on both the fixed and growing domains. 
 
Our approach enables us to model subdiffusive transport of particles on arbitrarily growing domains, and the solution of the auxiliary master equation on the fixed domain could be used as the basis for numerical simulations of subdiffusive transport on growing domains. The equations we derive on the growing domain can be interpreted phenomenologically as a reaction sub-diffusion process with an additional advective term. In this context, the reaction represents the dilution of the concentration due to the growing domain.

\begin{figure}[htbp]
\begin{center}
\includegraphics[width=0.4 \textwidth]{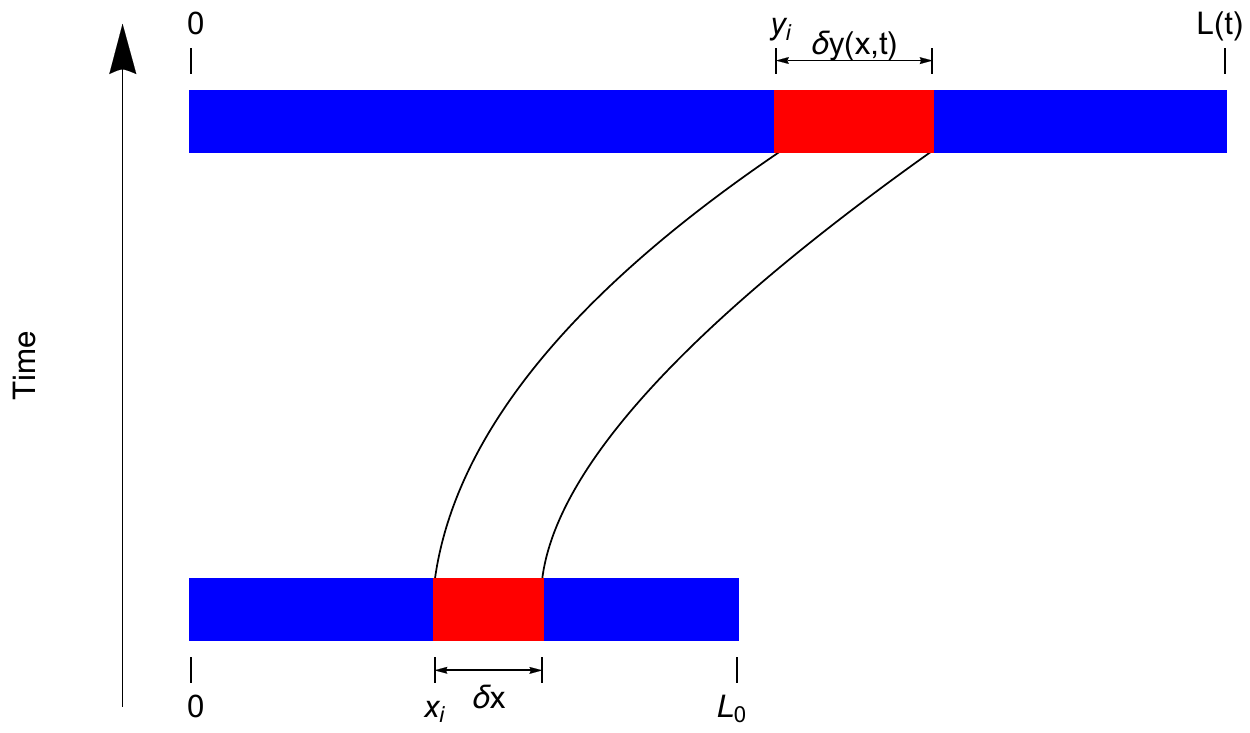} 
\caption{Schematic representation of the growth of the domain and the mapping of an interval in the initial domain to a corresponding interval at some later time $t$.}
\label{fig_map}
\end{center}
\end{figure}

We wish to construct a mapping between a location on the initial fixed domain, $x\in[0,L_0]$, to the corresponding location at some later time $t$, $y\in [0,L(t)]$. To characterise how the domain is changing in time we begin by partitioning the domain $[0,L_0]$ into $m$ cells of width $\delta x=\frac{1}{m}$. We will denote the boundary positions of this partition such that $x_i=i \delta x$.  As the domain grows, the width of the partitions, now denoted by $\delta y_i(t)$ will have grown with the domain and formed a partition of $[0,L(t)]$. Note that whilst the initial cell widths were constant this is no longer the case in the growing domain, i.e. $\delta y_i$ is a function of both the initial position $x_i$ and time. The mapping is defined through a growth function, $\mu(x_i,t)$, which defines the growth rate of the interval at $x_i$ at a time $t$. Explicitly it can be shown that the mapping $g(x,t)$ from a position in the fixed domain, $x$, to a corresponding position on the growing domain, $y$, is given by,
\begin{align}\label{eq_xymap}
y=\lim_{n\to\infty}\sum_{i=1}^n \delta y_i
=\int_{0}^{x}\exp\left(\int_0^{t}\mu(z,s)ds\right)dz=g(x,t).
\end{align}
This is illustrated schematically in Figure \ref{fig_map}. Note that, $g(0,t)=0$ and the initial condition, $y=g(x,0)=x$ for all $x\in [0,L_0]$, places a physical restriction on the mapping between $y$ and $x$. For future notational convenience we will denote the spatial derivative of $g(x,t)$ as $\nu^*(x,t)$, so that,
\begin{equation}
\nu^*(x,t)=\frac{\partial g(x,t)}{\partial x}=e^{\int_0^t \mu(x,s)ds},
\end{equation}
and the time derivative as,
\begin{equation}
\eta^*(x,t)=\frac{\partial g(x,t)}{\partial t}=\int_0^x \mu(z,t) e^{\int_0^t \mu(z,s)ds}dz.
\end{equation}
As the mapping is invertible, so that $x=g^{-1}(y,t)$, these can be expressed on the growing domain, giving,
\begin{equation}\label{eq_nu_y}
\nu(y,t)=\nu^*(g^{-1}(y,t),t),
\end{equation}
and
\begin{equation}\label{eq_eta_y}
\eta(y,t)=\eta^*(g^{-1}(y,t),t).
\end{equation}

It should also be noted that if we consider the growth of a small interval in the initial domain $(x,x+\delta x)$, then the width of the interval at some later time, in the limit of small $\delta x$ can be written as, 
\begin{equation}
\label{eq_dely_def}
\delta y(x,t)= e^{\int_0^t \mu(x,s)ds} \delta x.
\end{equation}

 We now consider a CTRW on a growing domain, such that a particle will jump to a location, wait for some time, and then jump to a new location. We will assume that the waiting time and jump length densities are independent. The waiting time probability density for a particle that arrived at a location at time $t'$ to jump at time $t$ will be denoted by $\psi(t-t')$, where $t-t'$ is the amount of time that the particle waited. The jump length density for a particle that is at a location $z'$ to jump to location $z$ at time $t$ is denoted by $\lambda(z|z',t)$. In the following we consider a CTRW on the growing domain $z=y$ and an auxiliary CTRW on the fixed domain  $z=x$. In taking the diffusion limit we will restrict ourselves to fixed length jumps on the growing domain, $\Delta y$, where the particle may jump either left or right. The corresponding jumps in the auxiliary CTRW on the initial fixed domain will therefore have lengths that change in both time and space as the domain grows.

For a particle undergoing a CTRW on the growing domain, we let $\rho(y,t)\delta y(x,t)$ denote the probability of finding the particle in the region $(y,y+\delta y(x,t))$, in the time $(t,t+\delta t)$ for a small $\delta y(x,t)$. Thus $\rho(y,t)$ is the probability density of finding the particle, which we can express as follows,
\begin{equation} \label{eq_rho0}
\rho(g(x,t),t)\delta y(x,t) =\int_{0}^{t}\Phi(t-t')q(g(x,t'),t')\delta y(x,t') dt',
\end{equation}
where $\Phi(t-t')$ is the survival function associated with the waiting time density $\psi(t-t')$.
The inbound flux, $q(g(x,t),t)$, is defined such that the probability of the particle entering the region $(y,y+\delta y(x,t))$ in the time $(t,t+\delta t)$, given $y=g(x,t)$, is $q(g(x,t),t)\delta y(x,t)\delta t$.
This equation states that for a particle to be in the region, it must have previously arrived in the region and not jumped away. 

Equation \eqref{eq_rho0} can be simplified by using Eq. \eqref{eq_dely_def},
\begin{equation} \label{eq_rho}
\begin{split}
\rho(g(x,t),t)e^{\int_0^t \mu(x,s)ds}&=\int_{0}^{t}\Phi(t-t')q(g(x,t'),t')e^{\int_0^{t'} \mu(x,s) ds} dt'.
\end{split}
\end{equation}  
To transform the evolution equation to a master equation it is necessary to replace the explicit dependence on $q(g(x,t),t)$ with a dependence on $\rho(g(x,t,),t)$. The growth of the domain requires us to utilise non-standard techniques to achieve this. As the region is moving and growing this is most easily expressed by mapping the required functions back to the fixed $x$ domain. 
The formulation of the CTRW on the fixed domain will be referred to as an auxiliary CTRW.

To formulate the auxiliary CTRW on the fixed domain, we relate the associated densities to densities on the growing domain, such that,
\begin{equation}
\rho(y,t)=\rho(g(x,t),t)=\rho^*(x,t),\;\;\;\;\;\;\;q(g(x,t),t)=q^*(x,t).
\end{equation}
Here we use a star to denote a function associated with the auxiliary process on the fixed domain. Hence we can write the auxiliary form of Eq. \eqref{eq_rho} as, 
\begin{equation}\label{eq_rho_star}
\rho^*(x,t)e^{\int_0^t \mu(x,s)ds}=\int_{0}^{t}\Phi(t-t')q^*(x,t')e^{\int_0^{t'} \mu(x,s) ds} dt'.
\end{equation}
Note that this left hand side, $\rho^*(x,t)e^{\int_0^t \mu(x,s)ds}$, is a conserved probability density. Differentiating Eq. \eqref{eq_rho_star} with respect to time and simplifying, we arrive at an evolution equation for the probability density,
\begin{equation} \label{eq_evo}
\frac{\partial \rho^*(x,t)}{\partial t}=q^*(x,t)-\int_0^t\psi(t-t')e^{-\int_{t'}^{t} \mu(x,s) ds}q^*(x,t')dt'-\mu(x,t)\rho^*(x,t).
\end{equation}
In this equation the second term on the right hand side is the flux out of the neighbourhood around $x$ in the time interval around $t$, while the third term is the reduction in concentration of particles, around $x$ around $t$, due to the growth of the domain. Explicitly we define the flux out as,
\begin{equation} \label{eq_out}
i^*(x,t)=\int_0^t\psi(t-t')q^*(x,t')e^{-\int_{t'}^{t} \mu(x,s) ds}dt'.
\end{equation}
In this equation the incoming flux, $q^*(x,t)$, can itself be expressed in terms of the flux out resulting in the relation,
\begin{equation}\label{eq_fluxin_from_fluxout}
\begin{split}
q^*(x,t)&=\int_0^{L(0)} \lambda(x|x',t)i^*(x',t)dx'.
\end{split}
\end{equation}

Using Eq. \eqref{eq_fluxin_from_fluxout}, noting the semi-group property of the exponential function, we can rewrite Eq. \eqref{eq_evo} and using Laplace transform methods, we can express the evolution equation for the auxiliary CTRW as the auxiliary master equation,
\begin{equation}
\label{eq_master}
\begin{split}
\frac{\partial \rho^*(x,t)}{\partial t}&=\int_0^{L(0)} \lambda(x|x',t) \int_0^t K(t-t')\rho^*(x',t')e^{-\int_{t'}^{t}\mu(x',s)ds}dt'dx'\\&-\int_0^t K(t-t')\rho^*(x,t')e^{-\int_{t'}^{t}\mu(x,s)ds}dt'-\mu(x,t)\rho^*(x,t).
\end{split}
\end{equation}
In this equation, the memory kernel, $K(t)$, is defined by,
\begin{equation} \label{eq_mem_kern}
\mathcal{L}_t\{K(t)\}=\frac{\mathcal{L}_t\{\psi(t)\}}{\mathcal{L}_t\{\Phi(t)\}},
\end{equation}
where $\mathcal{L}_t$ denotes a Laplace transform with respect to time.

The master equation, Eq. \eqref{eq_master}, has been derived for arbitrary waiting time and jump densities. 
To obtain a diffusion limit of the master equation we will require specific forms for these densities. We wish to consider the case of a fixed jump length on the growing domain, where the particle will jump either right or left with equal probability. In this case the jump length for the auxiliary master equation will change with both space and time. The jump probability density can therefore be written as,
\begin{equation}\label{eq_jump_onx}
\begin{split}
\lambda(x|x',t)&=\frac{1}{2}\left(\delta(x-x'-\epsilon^+)+\delta(x-x'+\epsilon^-)\right),
\end{split}
\end{equation}
where $\delta(x)$ is the Dirac delta function, and $\epsilon^+$ and $\epsilon^-$ are time and space dependent. To relate the $\epsilon$'s to the fixed jump length, $\Delta y$, we note that from Eq. (\ref{eq_xymap}) we have,
\begin{align}
\Delta y&=\int_{x-\epsilon^{+}}^{x}e^{\int_{0}^{t}\mu(z,s)ds}dz, \label{eq_eps+} \\
\Delta y&=\int_{x}^{x+\epsilon^{-}}e^{\int_{0}^{t}\mu(z,s)ds}dz. \label{eq_eps-}
\end{align}

Using the relations from Eqs. (\ref{eq_eps+}) and (\ref{eq_eps-}), we perform a Taylor expansion of Eq. \eqref{eq_master} with the jump distribution given by, Eq. \eqref{eq_jump_onx} around $\Delta y=0$ to arrive at,
\begin{equation} \label{eq_amq_tay}
\begin{split}
&\frac{\partial \rho^*(x,t)}{\partial t}= \frac{\Delta y^2e^{-2\int_{0}^{t}\mu(x,s)ds}}{2}\left(\left(\frac{\partial^2}{\partial x^2}\int_0^t K(t-t')\rho^*(x,t')e^{-\int_{t'}^t\mu(x,s)ds}dt'\right)\right.\\
&\left.-\left(\int_{0}^{t}\frac{\partial \mu(x,s)}{\partial x}ds\right)\left(\frac{\partial}{\partial x}\int_0^t K(t-t')\rho^*(x,t')e^{-\int_{t'}^t\mu(x,s)ds}dt'\right)\right)-\mu(x,t)\rho^*(x,t)+O(\Delta y^3).
\end{split}
\end{equation}

We now take a Mittag-Leffler waiting time density, given by,
\begin{equation} \label{eq:ML_waiting}
 \psi(t) =\frac{t^{\alpha-1}}{\tau^{\alpha}}E_{\alpha,\alpha} \left( -\left(\frac{t}{\tau}\right)^{\alpha} \right),
\end{equation}
with $0<\alpha<1$ and $\tau >0$ \cite{HA1995}, where
$E_{\alpha,\beta}$ is a two parameter Mittag-Leffler function defined by,
\begin{equation}
E_{\alpha,\beta}(z)=\sum_{k=0}^{\infty}\frac{z^{k}}{\Gamma(\alpha k+\beta)}.
\end{equation}
The Mittag-Leffler probability density is heavy tailed, which is asymptotically $\psi(t)\sim t^{-1-\alpha}$ for long times.
The memory kernel of a Mittag-Leffler probability density can be calculated from the inverse Laplace transform of Eq. (\ref{eq_mem_kern}),
\begin{equation}\label{eq_mem_ML}
K(t)=\mathcal{L}_s^{-1}\left\{\frac{s^{1-\alpha}}{\tau^{\alpha}}\right\}.
\end{equation}

The Riemann-Liouville fractional derivative of order $1-\alpha$ is defined as,
\begin{equation}
_0\mathcal{D}_{t}^{1-\alpha}\left(f(t)\right)=\frac{1}{\Gamma(\alpha)}\frac{d}{dt}\int_0^t f(t')(t-t')^{\alpha-1}dt'.
\end{equation}
As we are considering smooth real valued functions, the initial condition term in the Laplace transform of the Riemann-Liouville fractional derivative will be zero \cite{LD2007}, so that, 
\begin{equation}
\mathcal{L}_t\left\{_0\mathcal{D}_{t}^{1-\alpha}\left(f(t)\right)\right\}=s^{1-\alpha}\mathcal{L}_t\left\{f(t)\right\}.
\end{equation}

Using Mittag-Leffler distributed waiting times the auxiliary master equation on the fixed domain, Eq. (\ref{eq_amq_tay}), becomes,
\begin{equation} \label{eq_mlme}
\begin{split}
\frac{\partial \rho^*(x,t)}{\partial t}
=&\frac{\Delta y^2e^{-2\int_{0}^{t}\mu(x,s)ds}}{2\tau^\alpha}\left(\frac{\partial^2}{\partial x^2}\left(\frac{\,_0\mathcal{D}_t^{1-\alpha}\left(\rho^*(x,t)e^{\int_{0}^{t}\mu(x,s)ds}\right)}{e^{\int_{0}^{t}\mu(x,s)ds}}\right)\right.\\
&\left.-\left(\int_{0}^{t}\frac{\partial \mu(x,s)}{\partial x}ds\right)\frac{\partial}{\partial x}\left(\frac{\,_0\mathcal{D}_t^{1-\alpha}\left(\rho^*(x,t)e^{\int_{0}^{t}\mu(x,s)ds}\right)}{e^{\int_{0}^{t}\mu(x,s)ds}}\right)\right)-\mu(x,t)\rho^*(x,t)+\mathcal{O}(\Delta y^3).
\end{split}
\end{equation} 
The fractional diffusion limit is one in which the length and time scales are taken to zero, such that,
\begin{equation}\label{eq_anom_dif_lim}
D_\alpha=\lim_{\Delta y,\tau\to0}\frac{\Delta y^2}{2{\tau^\alpha}},
\end{equation}
exists. The fractional diffusion limit of Eq. (\ref{eq_mlme}) is,
\begin{equation}
\begin{split}
\label{eq_master_anom}
\frac{\partial \rho^*(x,t)}{\partial t}&=D_\alpha e^{-2\int_{0}^{t}\mu(x,s)ds}\left(\frac{\partial^2}{\partial x^2}\left(\frac{\,_0\mathcal{D}_t^{1-\alpha}\left(\rho^*(x,t)e^{\int_{0}^{t}\mu(x,s)ds}\right)}{e^{\int_{0}^{t}\mu(x,s)ds}}\right)\right.\\
&\left.-\left(\int_{0}^{t}\frac{\partial \mu(x,s)}{\partial x}ds\right)\frac{\partial}{\partial x}\left(\frac{\,_0\mathcal{D}_t^{1-\alpha}\left(\rho^*(x,t)e^{\int_{0}^{t}\mu(x,s)ds}\right)}{e^{\int_{0}^{t}\mu(x,s)ds}}\right)\right)-\mu(x,t)\rho^*(x,t).
\end{split}
\end{equation}
This is the auxiliary fractional diffusion equation defined on the fixed domain. Note that, apart from the advective type term, this is the same form as a fractional reaction subdiffusion equation \cite{ADH2013mmnp}, with the additional feature of a space and time dependent diffusivity. In writing the equation in terms of the growing domain coordinates the diffusivity will be constant.

Boundary conditions may be implemented by considering different jump length densities near the boundary. Explicitly, a zero flux boundary will be implemented by taking,
\begin{equation}
\label{eq_bound_r}
\lambda(x|x',t)=\delta(x-x'+\epsilon^-),
\end{equation}
for $x\in [L(0)-\epsilon^-, L(0)]$ and,
\begin{equation}
\label{eq_bound_l}
\lambda(x|x',t)=\delta(x-x'-\epsilon^+),
\end{equation}
for $x\in[0,\epsilon^{+}]$. This jump density guarantees that there is no flux across the boundary, and in the diffusive limit the master equation at the boundary point will be consistent with the master equation in the bulk. 

Using the jump length density for the left boundary, Eq. \eqref{eq_bound_l}, and taking a Taylor expansion around $\Delta y=0$, the master equation, Eq. \eqref{eq_master}, becomes,
\begin{equation} \label{eq_master_with_jump_taylor_bound}
\begin{split}
\frac{\partial \rho^*(x,t)}{\partial t}&=\Delta y e^{-\int_0^t \mu(x,s)ds}\frac{\partial}{\partial x}\left(\int_0^t K(t-t')\rho^*(x,t')e^{-\int_{t'}^t\mu(x,s)ds}dt'\right)\\& 
+\frac{\Delta y^2e^{-2\int_{0}^{t}\mu(x,s)ds}}{2}\left(\left(\frac{\partial^2}{\partial x^2}\int_0^t K(t-t')\rho^*(x,t')e^{-\int_{t'}^t\mu(x,s)ds}dt'\right)\right.\\
&\left.-\left(\int_{0}^{t}\frac{\partial \mu(x,s)}{\partial x}ds\right)\left(\frac{\partial}{\partial x}\int_0^t K(t-t')\rho^*(x,t')e^{-\int_{t'}^t\mu(x,s)ds}dt'\right)\right)-\mu(x,t)\rho^*(x,t)+O(\Delta y^3).
\end{split}
\end{equation}
for $x\in[0,\epsilon^{+}] $.
The difference between this equation and the bulk result is the occurrence of a first order spatial derivative. With the Mittag-Leffler waiting time density in order for the diffusion limit, Eq. \eqref{eq_anom_dif_lim} to exist, we require the first order spatial derivative term to be,
\begin{equation} \label{eq_bound_x}
\left.\frac{\partial}{\partial x}\left(\frac{\,_0\mathcal{D}_t^{1-\alpha}\left(\rho^*(x,t)e^{\int_{0}^{t}\mu(x,s)ds}\right)}{e^{\int_{0}^{t}\mu(x,s)ds}}\right)\right|_{x=0}=0.
\end{equation}
Only holding at the boundary point as $\Delta y \to 0$.  This zero flux boundary condition is equivalent to the zero flux boundary derived for fractional reaction subdiffusion equations \cite{ADHJL2016}. The derivation for the right hand side of the boundary results in an equivalent condition.

The fractional diffusion equation can be found by mapping the auxiliary equation, Eq. (\ref{eq_master_anom}), to the growing domain. Using the mapping $y=g(x,t)$, with  Eqs. \eqref{eq_nu_y} and \eqref{eq_eta_y}, we perform a change of variables and find,
\begin{equation}
\label{eq_master_anom_gd}
\begin{split}
\frac{\partial \rho(y,t)}{\partial t}+\eta(y,t)\frac{\partial \rho(y,t)}{\partial y}&=D_\alpha\frac{\partial^2}{\partial y^2}\left(\frac{1}{\nu(y,t)}\,_0^{g}\mathcal{C}_t^{1-\alpha}\left(\rho(y,t)\nu(y,t)\right)\right)\\
&-\left(\frac{\partial \nu(y,t)}{\partial t}\right)\frac{1}{\nu(y,t)}\rho(y,t).
\end{split}
\end{equation}
Here we have defined a new comoving fractional derivative, $\,_0^{g}\mathcal{C}_t^{1-\alpha}$, which operates along the curve, $y=g(x,t)$, for a fixed $x$. Formally this is defined as,
\begin{equation} \label{eq_new_fd}
\,_0^{g}\mathcal{C}_t^{1-\alpha}f(y,t)=\frac{1}{\Gamma(\alpha)}\frac{\partial}{\partial t}\int_0^t f(g(g^{-1}(y,t),t'),t')(t-t')^{\alpha-1}dt'.
\end{equation}
Informally, the history of the function is not integrated over a fixed value of $y$ but rather along the trajectory of the point in the domain as it grows. We note that, 
\begin{equation}
\,_0^{g}\mathcal{C}_t^{1-\alpha}\left(\rho(y,t)\nu(y,t)\right)=\,_0\mathcal{D}_t^{1-\alpha}\left(\rho^*(x,t)\nu^*(x,t)\right).
\end{equation}
The boundary condition, Eq. (\ref{eq_bound_x}), on the growing domain is,
\begin{equation}\label{eq_bound_ML_g}
\left.\frac{\partial}{\partial y}\left(\frac{\,^g_0\mathcal{C}_t^{1-\alpha}\left(\rho(y,t)\nu(y,t)\right)}{\nu(y,t)}\right)\right|_{y=0}=0.
\end{equation} 
We note that when the rate of the domain growth has no spatial dependence, i.e. $\nu(y,t)=f(t)$, the boundary conditions for the Mittag-Leffler waiting time density is simplified to,
\begin{equation}
\left.\frac{\partial\rho(y,t)}{\partial y}\right|_{y=0,L(t)}=0,
\end{equation}
on the growing domain.

In this work we have derived evolution equations that describe subdiffusive transport on a growing domain. Equation (\ref{eq_master_anom}) describes the transport on a rescaled fixed domain whilst Eq. (\ref{eq_master_anom_gd}) describes the same process on the growing domain. The evolution equation on the growing domain required the definition of a new fractional order differential operator that follows the domain growth, Eq. (\ref{eq_new_fd}). Our work provides a sound basis for further explorations of models of subdiffusive transport on a growing domain, including reactions, pattern formation, and morphogenesis.

This work was supported by the Australian Commonwealth Government (ARC DP140101193). We gratefully acknowledge useful discussions with S. B. Yuste.

\bibliographystyle{elsarticle-num}

\begin{thebibliography}{10}
\expandafter\ifx\csname url\endcsname\relax
  \def\url#1{\texttt{#1}}\fi
\expandafter\ifx\csname urlprefix\endcsname\relax\def\urlprefix{URL }\fi
\expandafter\ifx\csname href\endcsname\relax
  \def\href#1#2{#2} \def\path#1{#1}\fi

\bibitem{CGK1999}
E.~J. Crampin, E.~A. Gaffney, P.~K. Maini, Reaction and diffusion on growing
  domains: scenarios for robust pattern formation, Bull. Math. Biol. 61~(6)
  (1999) 1093--1120.
\newblock \href {http://dx.doi.org/10.1006/bulm.1999.0131}
  {\path{doi:10.1006/bulm.1999.0131}}.

\bibitem{M2001}
J.~D. Murray, Mathematical Biology. II Spatial Models and Biomedical
  Applications {Interdisciplinary Applied Mathematics V. 18}, Springer-Verlag
  New York Incorporated, 2001.

\bibitem{CM2001}
E.~Crampin, P.~Maini, Modelling biological pattern formation: the role of
  domain growth, Comments Theor. Biol. 6~(3) (2001) 229--249.

\bibitem{BYE2010}
R.~E. Baker, C.~A. Yates, R.~Erban, From microscopic to macroscopic
  descriptions of cell migration on growing domains, Bull. Math. Biol. 72~(3)
  (2010) 719--762.
\newblock \href {http://dx.doi.org/10.1007/s11538-009-9467-x}
  {\path{doi:10.1007/s11538-009-9467-x}}.

\bibitem{WBGM2011}
T.~E. Woolley, R.~E. Baker, E.~A. Gaffney, P.~K. Maini, Stochastic reaction and
  diffusion on growing domains: understanding the breakdown of robust pattern
  formation, Phys. Rev. E 84~(4) (2011) 046216.
\newblock \href {http://dx.doi.org/10.1103/PhysRevE.84.046216}
  {\path{doi:10.1103/PhysRevE.84.046216}}.

\bibitem{YBEM2012}
C.~A. Yates, R.~E. Baker, R.~Erban, P.~K. Maini, Going from microscopic to
  macroscopic on nonuniform growing domains, Phys. Rev. E 86~(2) (2012) 021921.
\newblock \href {http://dx.doi.org/10.1103/PhysRevE.86.021921}
  {\path{doi:10.1103/PhysRevE.86.021921}}.

\bibitem{SSMB2015}
M.~J. Simpson, J.~A. Sharp, L.~C. Morrow, R.~E. Baker, Exact solutions of
  coupled multispecies linear reaction--diffusion equations on a uniformly
  growing domain, PloS one 10~(9) (2015) e0138894.
\newblock \href {http://dx.doi.org/10.1371/journal.pone.0138894}
  {\path{doi:10.1371/journal.pone.0138894}}.

\bibitem{YAE2016}
S.~Yuste, E.~Abad, C.~Escudero, Diffusion in an expanding medium: Fokker-planck
  equation, green's function, and first-passage properties, Phys. Rev. E 94~(3)
  (2016) 032118.
\newblock \href {http://dx.doi.org/10.1103/PhysRevE.94.032118}
  {\path{doi:10.1103/PhysRevE.94.032118}}.

\bibitem{KA1995}
S.~Kondo, R.~Asai, A reaction-diffusion wave on the skin of the marine
  angelfish pomacanthus, Nature 376~(6543) (1995) 765.
\newblock \href {http://dx.doi.org/10.1038/376765a0}
  {\path{doi:10.1038/376765a0}}.

\bibitem{S2007}
M.~J. Saxton, A biological interpretation of transient anomalous subdiffusion.
  {I}. {Q}ualitative model, Biophys. J. 92~(4) (2007) 1178--1191.
\newblock \href {http://dx.doi.org/10.1529/biophysj.106.092619}
  {\path{doi:10.1529/biophysj.106.092619}}.

\bibitem{SWDA2006}
F.~Santamaria, S.~Wils, E.~{De Schutter}, G.~J. Augustine, Anomalous diffusion
  in purkinje cell dendrites caused by spines, Neuron 52~(4) (2006) 635--648.
\newblock \href {http://dx.doi.org/10.1016/j.neuron.2006.10.025}
  {\path{doi:10.1016/j.neuron.2006.10.025}}.

\bibitem{MNHMT2010}
G.~I. Mashanov, M.~Nobles, S.~C. Harmer, J.~E. Molloy, A.~Tinker, Direct
  observation of individual {KCNQ1} potassium channels reveals their
  distinctive diffusive behavior, J Biol Chem 285~(6) (2010) 3664--3675.
\newblock \href {http://dx.doi.org/10.1074/jbc.M109.039974}
  {\path{doi:10.1074/jbc.M109.039974}}.

\bibitem{WSTK2011}
A.~V. Weigel, B.~Simon, M.~M. Tamkun, D.~Krapf, Ergodic and nonergodic
  processes coexist in the plasma membrane as observed by single-molecule
  tracking, Proc. Natl. Acad. Sci. U.S.A. 108~(16) (2011) 6438--6443.
\newblock \href {http://dx.doi.org/10.1073/pnas.1016325108}
  {\path{doi:10.1073/pnas.1016325108}}.

\bibitem{SACSA2013}
S.~A. Shukair, S.~A. Allen, G.~C. Cianci, D.~J. Stieh, M.~R. Anderson, S.~M.
  Baig, C.~J. Gioia, E.~J. Spongberg, S.~M. Kauffman, M.~D. McRaven, et~al.,
  Human cervicovaginal mucus contains an activity that hinders {HIV}-1
  movement, Mucosal Immunol. 6~(2) (2013) 427--434.
\newblock \href {http://dx.doi.org/10.1038/mi.2012.87}
  {\path{doi:10.1038/mi.2012.87}}.

\bibitem{S2005}
A.~Shalchi, Time-dependent transport and subdiffusion of cosmic rays, Journal
  of Geophysical Research: Space Physics 110~(A9), a09103.
\newblock \href {http://dx.doi.org/10.1029/2005JA011214}
  {\path{doi:10.1029/2005JA011214}}.

\bibitem{LB2003}
M.~Levy, B.~Berkowitz, Measurement and analysis of non-{F}ickian dispersion in
  heterogeneous porous media, J Contam. Hydrol. 64~(3) (2003) 203 -- 226.
\newblock \href {http://dx.doi.org/10.1016/S0169-7722(02)00204-8}
  {\path{doi:10.1016/S0169-7722(02)00204-8}}.

\bibitem{AS2017}
S.~Abe, N.~Suzuki, Subdiffusion of volcanic earthquakes, Acta Geophysica (2017)
  1--9\href {http://dx.doi.org/10.1007/s11600-017-0029-6}
  {\path{doi:10.1007/s11600-017-0029-6}}.

\bibitem{HLW2006}
B.~I. Henry, T.~A.~M. Langlands, S.~L. Wearne, Anomalous diffusion with linear
  reaction dynamics: From continuous time random walks to fractional
  reaction-diffusion equations, Phys. Rev. E 74~(3) (2006) 031116.
\newblock \href {http://dx.doi.org/10.1103/PhysRevE.74.031116}
  {\path{doi:10.1103/PhysRevE.74.031116}}.

\bibitem{SSS2006}
I.~M. Sokolov, M.~G.~W. Schmidt, F.~Sagu{\'e}s, Reaction-subdiffusion
  equations, Phys. Rev. E 73~(3) (2006) 031102.
\newblock \href {http://dx.doi.org/10.1103/PhysRevE.73.031102}
  {\path{doi:10.1103/PhysRevE.73.031102}}.

\bibitem{F2010}
S.~Fedotov, Non-{M}arkovian random walks and nonlinear reactions: Subdiffusion
  and propagating fronts, Phys. Rev. E 81~(1) (2010) 011117.
\newblock \href {http://dx.doi.org/10.1103/PhysRevE.81.011117}
  {\path{doi:10.1103/PhysRevE.81.011117}}.

\bibitem{AYL2010}
E.~Abad, S.~Yuste, K.~Lindenberg, Reaction-subdiffusion and
  reaction-superdiffusion equations for evanescent particles performing
  continuous-time random walks, Phys. Rev. E 81~(3) (2010) 031115.
\newblock \href {http://dx.doi.org/10.1103/PhysRevE.81.031115}
  {\path{doi:10.1103/PhysRevE.81.031115}}.

\bibitem{ADH2013mmnp}
C.~N. Angstmann, I.~C. Donnelly, B.~I. Henry, Continuous time random walks with
  reactions forcing and trapping, Math. Model. Nath. Pheno. 8~(2) (2013)
  17--27.
\newblock \href {http://dx.doi.org/10.1051/mmnp/20138202}
  {\path{doi:10.1051/mmnp/20138202}}.

\bibitem{BMK2000}
E.~Barkai, R.~Metzler, J.~Klafter, From continuous time random walks to the
  fractional {Fokker-Planck} equation, Phys. Rev. E 61~(1) (2000) 132.
\newblock \href {http://dx.doi.org/10.1103/PhysRevE.61.132}
  {\path{doi:10.1103/PhysRevE.61.132}}.

\bibitem{SK2006}
I.~M. Sokolov, J.~Klafter, Field-induced dispersion in subdiffusion, Phys. Rev.
  Lett. 97~(14) (2006) 140602.
\newblock \href {http://dx.doi.org/10.1103/PhysRevLett.97.140602}
  {\path{doi:10.1103/PhysRevLett.97.140602}}.

\bibitem{HLS2010}
B.~I. Henry, T.~A.~M. Langlands, P.~Straka, Fractional {F}okker-{P}lanck
  equations for subdiffusion with space- and time-dependent forces, Phys. Rev.
  Lett. 105~(17) (2010) 170602.
\newblock \href {http://dx.doi.org/10.1103/PhysRevLett.105.170602}
  {\path{doi:10.1103/PhysRevLett.105.170602}}.

\bibitem{MW1965}
E.~Montroll, G.~Weiss, Random walks on lattices {II}, J. Math. Phys. 6 (1965)
  167.
\newblock \href {http://dx.doi.org/10.1063/1.1704269}
  {\path{doi:10.1063/1.1704269}}.

\bibitem{HA1995}
R.~Hilfer, L.~Anton, Fractional master equations and fractal time random walks,
  Phys. Rev. E 51~(2) (1995) R848.
\newblock \href {http://dx.doi.org/10.1103/PhysRevE.51.R848}
  {\path{doi:10.1103/PhysRevE.51.R848}}.

\bibitem{MK2000}
R.~Metzler, J.~Klafter, The random walk's guide to anomalous diffusion: A
  fractional dynamics approach, Phys. Rep. 339 (2000) 1--77.
\newblock \href {http://dx.doi.org/10.1016/S0370-1573(00)00070-3}
  {\path{doi:10.1016/S0370-1573(00)00070-3}}.

\bibitem{LD2007}
C.~Li, W.~Deng, Remarks on fractional derivatives, Appl. Math. Comput. 187~(2)
  (2007) 777--784.
\newblock \href {http://dx.doi.org/10.1016/j.amc.2006.08.163}
  {\path{doi:10.1016/j.amc.2006.08.163}}.

\bibitem{ADHJL2016}
C.~N. Angstmann, I.~C. Donnelly, B.~I. Henry, B.~Jacobs, T.~A. Langlands, J.~A.
  Nichols, From stochastic processes to numerical methods: A new scheme for
  solving reaction subdiffusion fractional partial differential equations, J.
  Comput. Phys. 307 (2016) 508--534.
\newblock \href {http://dx.doi.org/10.1016/j.jcp.2015.11.053}
  {\path{doi:10.1016/j.jcp.2015.11.053}}.

\end{thebibliography}

\end{document}